\documentclass{article}

\PassOptionsToPackage{numbers, compress}{natbib}

\usepackage[final]{neurips_2021}

\usepackage[utf8]{inputenc} 
\usepackage[T1]{fontenc}    
\usepackage{hyperref}       
\usepackage{url}            
\usepackage{booktabs}       
\usepackage{amsfonts}       
\usepackage{nicefrac}       
\usepackage{microtype}      
\usepackage{xcolor}         
\usepackage{amsmath}         
\usepackage{amssymb}         
\usepackage{graphicx}

\title{Towards Universal Atomic Composability: A Formal Model for Multi-Rollup Environments on Ethereum}

\author{Dipankar Sarkar \\
  Cryptuon Research \\
  \texttt{me@dipankar.name} \\
}

\begin{document}

\maketitle

\begin{abstract}
In the rapidly evolving domain of distributed ledger technology, scalability and interoperability have become paramount challenges for both academic and industry sectors. In this paper, we introduce a comprehensive formal model to address atomic composability across multiple rollups on Ethereum. The proposed model incorporates mechanisms like buffering, dependency management, concurrency control, and the groundbreaking zero-knowledge proofs. Moreover, we evaluate its practical repercussions, strengths, and weaknesses, ensuring resilience against manipulative or erroneous actions. The application of the proposed model to shared sequencers and other existing solutions accentuates its versatility and universality.
\end{abstract}

\section{Introduction}\label{introduction}

In the rapidly evolving domain of distributed ledger technology, scalability and interoperability have become paramount challenges for both academic and industry sectors. Ethereum, recognized as a pioneering smart contract platform, has initiated a myriad of advancements, with rollups being a significant answer to the blockchain trilemma: balancing scalability, security, and decentralization \cite{buterin2020ethereum}. However, as rollups appear promising for scalability, they might unintentionally lead to fragmented composability. Given the intertwining nature of systems and applications, ensuring atomicity in transactions across systems is vital.

Atomic composability is predicated on the principle that a transaction (A) can only be finalized if another transaction (B) is likewise finalized \cite{micali2016atomic}. For decentralized applications that operate over multiple rollups, this assurance is critical. Yet, actualizing this atomicity with disconnected rollups on Ethereum presents major obstacles.

This paper offers a thorough formal model that addresses atomic composability across multiple rollups on Ethereum. Incorporating insights from established distributed system solutions and contemporary cryptographic methodologies, the proposed model encompasses buffering, dependency management, concurrency control, and the groundbreaking zero-knowledge proofs \cite{ben2013snarks}. Beyond proposing the model, we evaluate its practical repercussions, strengths, and weaknesses, ensuring resilience against manipulative or erroneous actions.

Our intent extends beyond presenting a solution; we seek to stimulate a wider discourse on the future trajectory of interconnected blockchains. With a surge in applications shifting to a multi-rollup framework on Ethereum and elsewhere, a robust system guaranteeing atomic composability becomes indispensable. Through our model and ensuing discussions, we aim to make substantial contributions to this burgeoning field of blockchain study.

To grasp the intricacies of composability between rollups on Ethereum, it is imperative to first delineate the nature of rollups and composability, before exploring the challenges in achieving cross-rollup atomic composability.

\subsection{What are Rollups?}\label{what-are-rollups}

Rollups, in the Ethereum context, are scaling mechanisms that bolster
network throughput. They operate by conducting transactions off-chain
and subsequently submitting a transaction summary to the primary
Ethereum chain, thus enhancing transaction capacity without overloading
the main Ethereum network \cite{buterin2020ethereum}.

\subsection{What is Composability?}\label{what-is-composability}

Within blockchain and Ethereum paradigms, composability pertains to the
capability of decentralized applications (dApps) and smart contracts to
effortlessly integrate and leverage one another's features \cite{schar2020decentralized}. This can be analogized to ``money Legos'', where each protocol or dApp represents an individual Lego piece, capable of diverse
combinations.

\subsection{Limits of Composability Between
Rollups}\label{limits-of-composability-between-rollups}

With the deployment of multiple rollups on Ethereum, challenges arise in
ensuring seamless interaction of dApps and contracts across these
rollups. This dilemma is intensified if rollups operate in isolation or
lack an effective bridging mechanism.

\textbf{Atomic Transactions Across Rollups:} Ensuring atomic
composability between rollups necessitates that a transaction in one
rollup is only finalized if its counterpart in another rollup is as
well. This is intricate because each rollup might possess unique
consensus algorithms, validation methodologies, and operational latency.

\textbf{Data Availability:} For a contract in one rollup to interface
with data or another contract on a distinct rollup, the requisite data
from the latter may not be readily accessible or may be costly to
retrieve.

\textbf{Differing Rules and Standards:} Distinct rollups with divergent
standards or rules regarding transaction processing can further impede
cross-rollup interactions.

\section{Formal Model - Rollups with Decentralized Common Pool
(DCP)}\label{formal-model-for-rollups-with-decentralized-common-pool-dcp}

Addressing the multifaceted nature of blockchain ecosystems, especially
those spanning several rollups, demands a structured, rigorous approach
to uphold transactional integrity and reliability. To this end, our
formal model seeks to methodically dissect and elucidate atomic
composability across Ethereum's multiple rollups.

Our model is an amalgamation of classic distributed system theories and
innovative cryptographic practices. We recognize that merely adapting
traditional system theories to the blockchain milieu is not sufficient,
necessitating a model tailored for the peculiarities of decentralized
ledgers, especially within Ethereum's ecosystem (\cite{narayanan2016bitcoin}.

Subsequent sections will delve into the intricacies of our formal model,
commencing with fundamental definitions. We will then probe into
operational dynamics, examining transactional workflows, dependency
resolutions, and concurrency nuances. Cryptographic methodologies,
particularly zero-knowledge proofs, will be highlighted, underscoring
their pivotal role in efficient, confidential transaction validations.

With this formal model, our aspiration is to furnish readers with an
encompassing, lucid, and rigorous comprehension of how to establish,
sustain, and, if necessary, re-establish atomic composability in
environments with multiple rollups.

\subsection{Definitions}

\begin{itemize}
    \item \( R \): Set of rollups on Ethereum.
    \item \( T \): Set of transactions, where \( T_{i,j} \) is the j-th transaction on rollup \( R_i \).
    \item \( P_d \): Decentralized common pool.
    \item \( K \): Set of cryptographic keys associated with transactions.
    \item \( \tau \): Timestamp attached to each transaction when it's accepted by the majority of \( P_d \) nodes.
    \item \( B \): Buffer zone where transactions with pending dependencies are stored.
    \item \( \tau_{\text{max}} \): Maximum time a transaction can reside in buffer \( B \).
    \item \( B_{\text{max}} \): Maximum number of transactions that can reside in buffer \( B \).
    \item \( D_{\text{max}} \): Maximum number of attempts to resolve a transaction's dependencies.
\end{itemize}

\subsection{Operations}

\subsubsection{Publish}
\begin{itemize}
    \item \textbf{Description}: This operation ensures that every transaction \( T_{i,j} \) from rollup \( R_i \) is published to the Decentralized Common Pool \( P_d \).
    \item \textbf{Steps}:
    \begin{enumerate}
        \item \( R_i \) generates a transaction \( T_{i,j} \).
        \item \( R_i \) sends \( T_{i,j} \) to \( P_d \).
        \item On receiving \( T_{i,j} \), the majority of nodes in \( P_d \) timestamp it with \( \tau \) and store it.
    \end{enumerate}
    \item \textbf{Output}: \( \text{publish}(T_{i,j}) \rightarrow P_d(\tau) \)
\end{itemize}

\subsubsection{Buffer}
\begin{itemize}
    \item \textbf{Description}: Transactions with unmet dependencies are sent to the buffer \( B \) until their dependencies are resolved or they hit one of the defined limits.
    \item \textbf{Steps}:
    \begin{enumerate}
        \item If \( T_{i,j} \) has unresolved dependencies, it’s directed to \( B \).
        \item \( T_{i,j} \) resides in \( B \) until either its dependencies are resolved or it breaches one of the constraints (\( \tau_{\text{max}} \), \( B_{\text{max}} \), or \( D_{\text{max}} \)).
    \end{enumerate}
    \item \textbf{Output}: \( \text{buffer}(T_{i,j}) \rightarrow B \)
\end{itemize}

\subsubsection{Resolve}
\begin{itemize}
    \item \textbf{Description}: The operation checks for and resolves dependencies between two transactions, possibly from different rollups.
    \item \textbf{Steps}:
    \begin{enumerate}
        \item Given two transactions \( T_{i,j} \) and \( T_{k,l} \), \( P_d \) checks for their mutual dependencies.
        \item If dependencies exist, \( P_d \) attempts resolution based on available data and timestamps.
        \item If resolution is successful within the constraints, the transactions proceed. Otherwise, they remain in \( B \) or are rejected.
    \end{enumerate}
    \item \textbf{Output}: \( \text{resolve}(T_{i,j}, T_{k,l}) \rightarrow \text{bool} \)
\end{itemize}

\subsubsection{Verify}
\begin{itemize}
    \item \textbf{Description}: Ensures the validity and authenticity of the transaction using cryptographic keys.
    \item \textbf{Steps}:
    \begin{enumerate}
        \item For each transaction \( T_{i,j} \), \( P_d \) uses the associated cryptographic key \( K_{i,j} \) to verify its authenticity and integrity.
        \item If the verification succeeds, the transaction proceeds. Otherwise, it’s deemed invalid.
    \end{enumerate}
    \item \textbf{Output}: \( \text{verify}(T_{i,j}, K_{i,j}) \rightarrow \text{bool} \)
\end{itemize}

\subsection{Transaction Processing}

For two transactions \( T_{i,j} \) from \( R_i \) and \( T_{k,l} \) from \( R_k \):

\subsubsection{Timestamp-Based Handling}
    If both transactions are timestamped in \( P_d \) and their timestamps are within an acceptable time difference (delta):
    \[ |\tau(T_{i,j}) - \tau(T_{k,l})| \leq \delta \]
    Then, the transactions are deemed compatible and can be processed without buffering.

\begin{figure}[ht]
  \centering
  \includegraphics[scale=0.4]{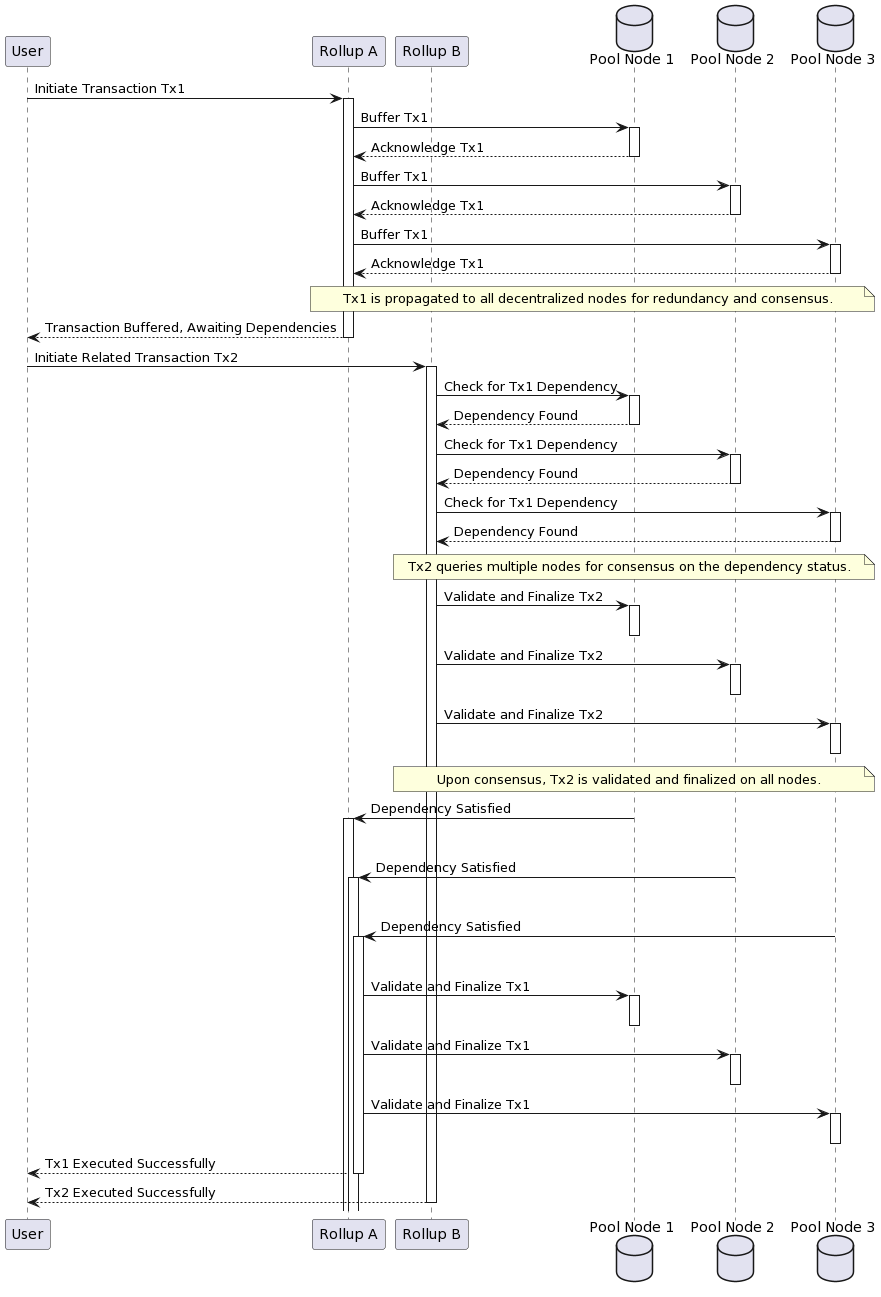}
  \caption{Sequence diagram for the formal model}
\end{figure}

\subsubsection{Buffering}
\begin{itemize}
    \item If the timestamp difference exceeds the acceptable delta, or if there’s a dependency which isn’t yet satisfied, transactions are sent to \( B \).
    \item While in \( B \):
    \begin{itemize}
        \item Periodic checks are done to see if dependencies can now be resolved.
        \item If \( (\tau_{\text{current}} - \tau(T_{i,j})) > \tau_{\text{max}} \), \( T_{i,j} \) is rejected.
        \item If the buffer size exceeds \( B_{\text{max}} \), a rejection policy \( P_R \) is triggered to create space.
        \item If attempts to resolve dependencies for \( T_{i,j} \) exceed \( D_{\text{max}} \), \( T_{i,j} \) is rejected.
    \end{itemize}
\end{itemize}

\subsubsection{Rejection and Notification}
Once a transaction is rejected, a notification mechanism informs the originating rollup \( R_i \) or the respective party about the rejection. This allows for potential re-submission or other actions from the user’s end.

\subsection{Rollup Punitive Measures}

\begin{figure}[h]
  \centering
  \includegraphics[scale=0.5]{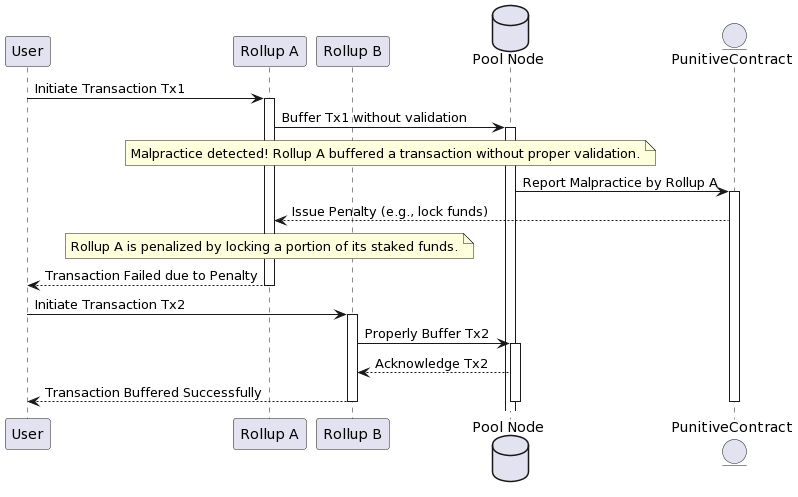}
  \caption{Sequence diagram for rollup misbehaviour}
\end{figure}

\subsubsection{Staking Mechanism}
Every rollup must stake a certain number of tokens to participate in the transaction composability system. This stake acts as collateral that can be forfeited or slashed if the rollup misbehaves.

\subsubsection{Monitoring and Reporting}
Third-party nodes or participants (often called "watchers" or "validators") can monitor transactions and execution behaviors across rollups. If they detect a rollup executing a transaction without proper validation or not adhering to the model's rules, they can submit a proof of this misbehavior.

\subsubsection{Misbehavior Proof}
A system can be put in place to accept proofs of misbehavior. Once verified, punitive measures can be enacted on the misbehaving rollup, including forfeiting their staked tokens.

\subsection{Decentralized Common Pool (DCP)}

Instead of having a centralized common pool, we can utilize a decentralized common pool, \( P_d \). This pool would be maintained and updated by a network of nodes, ensuring redundancy and security. Utilizing a consensus mechanism, these nodes can agree on the state of the pool.

\subsubsection{Core features}

\begin{itemize}
    \item \textbf{Consensus Mechanism}: Nodes in the \( P_d \) network use a consensus mechanism (like PoS, PoA, or a Byzantine Fault Tolerance mechanism) to validate and agree on the state of the pool. This prevents any single entity from maliciously modifying the data.
    
    \item \textbf{Data Redundancy}: The decentralized nature ensures that multiple copies of the transaction pool are stored across nodes. This redundancy makes it resistant to single points of failure and data tampering.
    
    \item \textbf{Cryptographic Verification}: Apart from using cryptographic keys for transaction verification, node participation, data propagation, and pool updates in the \( P_d \) network are also secured using cryptographic techniques, ensuring data integrity and authenticity.
\end{itemize}

\subsubsection{Concurrency Handling with Transaction Delay/Rejection}

\begin{itemize}
    \item \textbf{Timestamping}: Each transaction that's published to \( P_d \) is timestamped. This timestamp is based on when the transaction is received by the majority of nodes in the \( P_d \) network.
    
    \item \textbf{Transaction Buffer}: Transactions that have dependencies across rollups are not immediately processed. Instead, they're placed in a buffer zone.
    
    \item \textbf{Dependency Resolution with Buffer}: During the resolution phase, the system checks the buffer to resolve dependencies. If Transaction \( T_{i,j} \) depends on Transaction \( T_{k,l} \), the system will:
    \begin{itemize}
        \item Check if both transactions are in the buffer and if their timestamps are within an acceptable time difference (delta).
        \item If they are, the system processes both transactions.
        \item If not, and if the dependency can't be resolved within a certain timeframe, the system might either:
        \begin{enumerate}
            \item Reject one or both of the transactions.
            \item Delay the transaction processing until the dependency is resolved or a timeout occurs.
        \end{enumerate}
    \end{itemize}
\end{itemize}

\begin{figure}[h]
  \centering
  \includegraphics[scale=0.5]{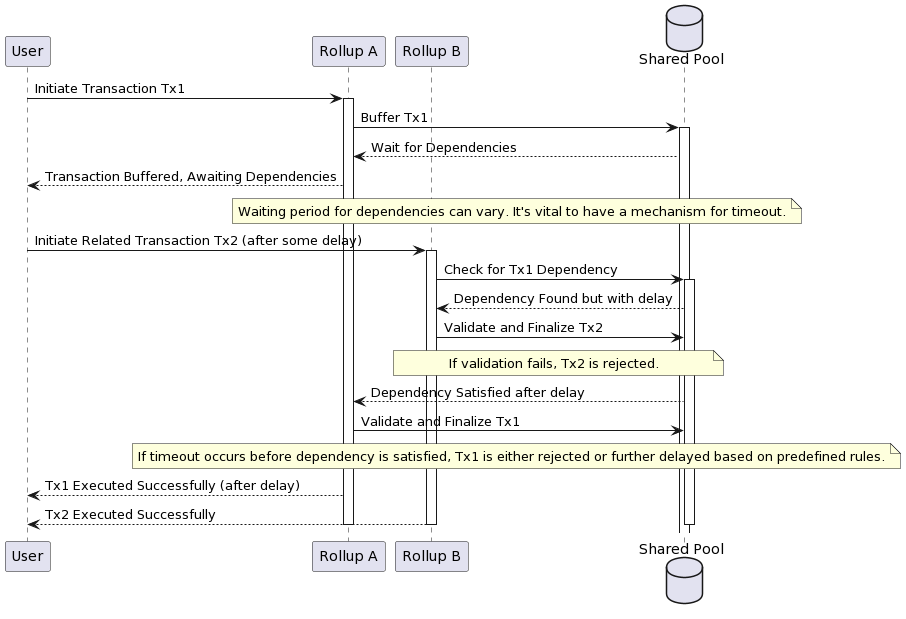}
  \caption{Sequence diagram with timeouts}
\end{figure}

\subsubsection{Challenges}

With the incorporation of a decentralized common pool, timestamping, a transaction buffer for dependency resolution, and enhanced security measures, the solution addresses many of the concurrency and security concerns. However, it's essential to consider trade-offs:

\begin{itemize}
    \item The system might have increased complexity due to decentralization and consensus mechanisms.
    \item Transaction delays, while managing concurrency, might not be suitable for applications that require real-time processing.
    \item Ensuring all nodes in the \( P_d \) network are updated in near real-time might present scalability challenges.
\end{itemize}
\section{Atomic composability \&
ZK-proofs}\label{atomic-composability-zk-proofs}

Zero-Knowledge Proofs (zk-proofs), particularly zk-SNARKs
(Zero-Knowledge Succinct Non-Interactive Argument of Knowledge) and
zk-STARKs (Zero-Knowledge Scalable Transparent Argument of Knowledge),
are cryptographic methods that allow one party to prove to another party
that a statement is true, without revealing any specific information
beyond the validity of the statement itself \cite{ben2013snarks, ben2018scalable}.

For atomic composability across rollups, zk-proofs can be exceptionally
beneficial. Let's explore how:

\subsection{Transaction Validation}\label{transaction-validation}

zk-proofs can be utilized to validate that a transaction on one rollup
adheres to specific conditions, without actually revealing the contents
of the transaction. This is especially useful for maintaining privacy
across rollups while still ensuring that conditions are met \cite{gennaro2013quadratic}.

\subsection{Dependency Verification}\label{dependency-verification}

If one transaction depends on another from a different rollup, zk-proofs
can be utilized to validate the successful execution and correctness of
the dependent transaction, again, without revealing the actual
transaction details \cite{micali1994cs}.

\subsection{Concurrency and Aggregate
Dependencies}\label{concurrency-and-aggregate-dependencies}

zk-proofs can be crafted to provide proofs of concurrent transaction
executions or aggregate transaction conditions (like total transaction
value across multiple rollups) being met, all without revealing specific
transaction details \cite{wahby2019fast}.

\subsection{Compactness and
Efficiency}\label{compactness-and-efficiency}

zk-proofs, especially zk-SNARKs, have the advantage of being succinct.
That means, irrespective of the amount of data or the number of
transactions they're validating, the proof size remains relatively small
and verification is swift. This feature can be immensely beneficial in a
system with multiple rollups, where swift validations are essential \cite{reitwiessner2016zk}.

\section{Incorporating Zk-proofs}\label{incorporating-zk-proofs}

To integrate zk-proofs into the model effectively, it would require the
rollups participating in this system to be zk-proof compatible. They
should be able to generate and verify these proofs efficiently.
Furthermore, standardization of proofs related to atomic composability
would be necessary to ensure smooth inter-rollup operations \cite{buterin2014next}.

\begin{enumerate}
\def\labelenumi{\arabic{enumi}.}
\item
  \textbf{Verification Functions}: Introduce verification functions
  within the model that use zk-proofs. These functions can quickly
  validate the correctness and completion of dependent transactions
  without needing full transparency into the transactions \cite{ben2018scalable}.
\item
  \textbf{Reduced Buffering Requirement}: With zk-proofs validating
  transaction dependencies almost immediately, the need for extensive
  buffering can be reduced. Transactions can be executed swiftly after
  their zk-proof verifications succeed \cite{bitansky2012extractable}.
\item
  \textbf{Privacy Maintenance}: As zk-proofs can validate statements
  without revealing the underlying data, transactions across rollups can
  maintain higher degrees of privacy, even in interdependent scenarios \cite{ben2013snarks}.
\item
  \textbf{Slashing with zk-proofs}: The monitoring and reporting
  mechanism can also utilize zk-proofs. Watchers can provide a zk-proof
  of misbehavior, which if validated, can lead to punitive measures \cite{wahby2019fast}.
\item
  \textbf{Proofs of Dependency Resolution}: In the case of cyclic or
  complex dependencies, zk-proofs can be crafted to ensure that all
  necessary conditions across rollups have been met without revealing
  the specifics of the transactions involved \cite{gennaro2013quadratic}.
\end{enumerate}

The integration of zk-proofs does introduce added cryptographic
complexity, but with the advantages of swift validation, reduced need
for buffering, and enhanced privacy, they can significantly bolster the
robustness and efficiency of the atomic composability model across
rollups \cite{williamson2020rollup}.

\section{Application of the Formal
Model}\label{application-of-the-formal-model}

As the Ethereum ecosystem evolves, diverse scaling solutions have
emerged to address the challenges of throughput and latency. Among them,
the concept of shared sequencers has gained significant traction. Shared
sequencers act as centralized transaction ordering mechanisms,
increasing throughput by temporarily assuming the role of transaction
orderer before these transactions are batched and finalized on the main
chain. While they introduce efficiencies, shared sequencers inherently
operate differently from typical rollups, warranting an exploration of
how our formal model for atomic composability can be applied to them.

The application of our formal model to shared sequencers and other
existing solutions accentuates its versatility and universality. Whether
it's the centralized nature of sequencers or the diverse architectures
of other rollups, the principles of atomic composability, as proposed in
our model, remain consistent. This consistency is instrumental in
creating a cohesive, interoperable, and scalable Ethereum ecosystem that
can cater to the ever-growing demands of decentralized applications and
services.

In this section, we extend our formal model to understand its
implications on shared sequencers and juxtapose it with other prevailing
solutions in the space. The goal is to provide a comparative analysis
that not only elucidates the strengths and potential drawbacks of each
approach but also offers a cohesive understanding of how atomic
composability can be universally achieved irrespective of the underlying
scaling solution.

\subsection{Shared Sequencers}\label{shared-sequencers-a-brief-overview}

Shared sequencers, by design, centralize the transaction ordering
mechanism without compromising the security guarantees of the main
chain. Transactions are quickly processed off-chain by the sequencer and
then aggregated into larger batches to be submitted on-chain. This
architecture provides the dual benefit of swift transaction times and
reduced on-chain congestion.

\subsubsection{Implications of the Formal Model:}\label{applying-the-formal-model}

\begin{enumerate}
\def\labelenumi{\arabic{enumi}.}
\item
  \textbf{Buffering \& Dependency Handling}: Given that shared
  sequencers operate in an off-chain environment before finalization,
  the buffering mechanism we proposed becomes even more critical. It
  allows for the temporary storage of transactions, especially when
  there are cross-rollup or cross-sequencer dependencies.
\item
  \textbf{Concurrency Control}: Shared sequencers inherently deal with a
  high volume of simultaneous transactions. Implementing our concurrency
  control mechanism ensures that interdependent transactions, even from
  different sources or rollups, can be processed in an atomic fashion.
\item
  \textbf{Zero-Knowledge Proofs \& Validation}: The validation mechanism
  using zk-proofs becomes an asset here. It ensures that even in a
  semi-centralized environment, transaction validations remain private,
  swift, and secure. The zk-proofs also provide an added layer of trust
  to users who might be skeptical of the centralized nature of
  sequencers.
\end{enumerate}

\subsection{Alternative solutions:}\label{comparison-with-existing-solutions}

While shared sequencers present a compelling case for scalability, other
solutions like zk-rollups, optimistic rollups, and sidechains each have
their unique architectures and merits.

\begin{enumerate}
\def\labelenumi{\arabic{enumi}.}
\item
  \textbf{zk-Rollups}: These rely heavily on zk-proofs for batched
  transaction validation. Applying our formal model, zk-rollups can
  benefit from enhanced dependency handling and buffering mechanisms,
  ensuring transactions across rollups are consistently processed.
\item
  \textbf{Optimistic Rollups}: Here, transactions are assumed to be
  correct until proven otherwise. Our formal model introduces a
  systematic approach for handling disputes and reordering, ensuring
  atomic composability without extensive delays.
\item
  \textbf{Sidechains}: Being independent blockchains, sidechains can
  pose more significant challenges for atomic composability. Our model
  can act as a bridge, providing mechanisms like dependency resolution
  and buffering to ensure smooth inter-chain operations.
\end{enumerate}

\section{Conclusion}\label{conclusion}

The formal model presented serves as a superset for all possible atomic
composability across rollups because of its comprehensive nature.

This model is a superset because it is designed to encompass every step,
every entity, and every possible scenario in the lifecycle of a
transaction across rollups. By being exhaustive in its approach, any
atomic composability solution across rollups that exists or might be
conceived in the future can be mapped onto some subset of this model. 

It acts as a general framework or blueprint from which specific
implementations can be derived, tailored to particular requirements or
constraints.

\textbf{Comprehensiveness of Definitions} It identifies every rollup, transaction, timestamp, cryptographic key,
buffer mechanism, and even the decentralized pool where transactions are
posted. This covers all essential tools and entities required for atomic
transactions between rollups.

\textbf{Detailed Operations} The detailed operation steps ensure that every conceivable action associated with transactions -- from their creation, publication,
verification, buffering, to dependency resolution -- is incorporated:
\begin{itemize}
\item
  \textbf{Publish}: Every transaction is committed to a decentralized
  common pool, capturing the universal broadcast mechanism.
\item
  \textbf{Buffer}: Handles the uncertainties and delays.
\item
  \textbf{Resolve}: Manages interdependencies, making sure that if one
  transaction in a set can't be executed, none in that set are.
\item
  \textbf{Verify}: Ensures that only legitimate and valid transactions
  are processed.
\end{itemize}

\textbf{Robust Dependency Handling} This is perhaps the crux of atomic composability. The model provides:

\begin{itemize}
\item
  A method to check the timing (timestamps) of transactions, which is
  crucial for ensuring order and dependencies.
\item
  Buffering mechanisms to account for delays in dependency resolution.
\item
  Defined limits for buffering to prevent infinite waits and to give an
  outcome (acceptance or rejection) within a finite time.
\end{itemize}

\textbf{Flexibility \& Scalability} The model doesn't restrict the number or type of dependencies. It simply
provides mechanisms to handle them. This ensures that as blockchain
technology evolves and new dependency types emerge or transactions
become more intricate, the model remains applicable.

\textbf{Incorporation of Limits} By integrating system limits like \( \tau_{\text{max}} \), \( B_{\text{max}} \), and \( D_{\text{max}} \), the model not only accounts for ideal scenarios
where all dependencies are quickly resolved but also for edge cases
where system constraints come into play. This adds to its universality.

\textbf{Security through Verification} The model's inclusion of cryptographic verification ensures that security concerns are front and center. It recognizes that composability isn't just about making sure transactions work together, but that they're also genuine and untampered.

\bibliographystyle{abbrvnat}
\bibliography{sample}
\end{document}